# Adsorption-controlled growth of MnTe(Bi$_2$Te$_3$)$_n$ by molecular beam epitaxy exhibiting stoichiometry-controlled magnetism

Jason Lapano[1], Lauren Nuckols[2], Alessandro R. Mazza[1], Yun-Yi Pai[1], Jie Zhang[1], Ben Lawrie[1], Rob G. Moore[1], Gyula Eres[1], Ho Nyung Lee[1], Mao-Hua Du[1], T. Zac Ward[1], Joon Sue Lee[3], William J. Weber[2], Yanwen Zhang[1,2], Matthew Brahlek[1]*

[1]Materials Science and Technology Division, Oak Ridge National Laboratory, Oak Ridge, TN, 37831, USA
[2]Department of Materials Science and Engineering, University of Tennessee, Knoxville, Tennessee 37996, USA
[3]Department of Physics and Astronomy, University of Tennessee, Knoxville, Tennessee 37996, USA

Correspondence should be addressed to *brahlekm@ornl.gov

**Abstract**: We report the growth of the intrinsic magnetic topological system MnTe(Bi$_2$Te$_3$)$_n$ by molecular beam epitaxy. By mapping the temperature and the Bi:Mn flux ratio, it is shown that there is a narrow growth window for the $n$=1 phase MnBi$_2$Te$_4$ with 2.0<Bi:Mn<2.6 at 225 °C. Here the films are stoichiometric and excess Bi and Te is not incorporated. At higher flux ratios (Bi:Mn≥4.5) it is found that the $n$ = 2 MnBi$_4$Te$_7$ phase is stabilized. Transport measurements indicate that the MnBi$_2$Te$_4$ and MnBi$_4$Te$_7$ undergo magnetic transitions around 25 K, and 10 K, respectively, consistent with antiferromagnetic phases found in the bulk. Further, for Mn-rich conditions (Bi:Mn<2), ferromagnetism emerges that exhibits a clear hysteretic state in the Hall effect, which likely indicates Mn-doped MnBi$_2$Te$_4$. Understanding how to grow ternary chalcogenide phases is the key to synthesizing new materials and to interface magnetism and topology, which together are routes to realize and control exotic quantum phenomena.

This manuscript has been authored by UT-Battelle, LLC under Contract No. DE-AC05-00OR22725 with the U.S. Department of Energy. The United States Government retains and the publisher, by accepting the article for publication, acknowledges that the United States Government retains a non-exclusive, paid-up, irrevocable, world-wide license to publish or reproduce the published form of this manuscript, or allow others to do so, for United States Government purposes. The Department of Energy will provide public access to these results of federally sponsored research in accordance with the DOE Public Access Plan (http://energy.gov/downloads/doe-public-access-plan).

Finding materials that exhibit multiple phenomena that are coupled gives deep insight into the physics of complex materials. Further, this can enable routes to design and engineer new technologies where modifying one state may induce switching in the other. Key examples are when a material is magnetic while concurrently (or incipiently) superconducting [1–5], ferroelectric [6–9] or more recently topological [10–14]. This latter example is of fundamental interest since magnetic ordering can be used to modify the topological class by changing the band structure [15–17]. Prominent examples are a Weyl semimetal phase that emerges from a 3D Dirac semimetal when ferromagnetism is introduced [18–21], as well as the quantized anomalous Hall (QAH) phase that emerges out of topological insulators when time reversal symmetry is broken [16,22–24]. From a materials viewpoint this can be achieved either through doping a topological material into a magnetic phase, interfacing it with a magnet as an atomic scale heterostructure, or by finding a material that intrinsically exhibits both phases. While magnetic doping has proven very effective at realizing the QAH in Cr/V-doped $(Bi_{1-x}Sb_x)_2Te_3$ [22,24], promising examples of the latter have recently emerged with the discovery of several classes of magnetic topological materials including Fe-Sn [25], $Co_3Sn_2S_2$ [18,21], and, of interest to the current work, the $MnTe(Bi_2Te_3)_n$ series [23,26–33].

The $MnTe(Bi_2Te_3)_n$ series is especially interesting due to the simultaneous topological nature along with the layered A-type antiferromagnetic (AFM) ground state, which, together, enabled the recent observation of the QAH in exfoliated bulk crystals of $MnBi_2Te_4$ [23,26]. As shown in Fig. 1, $MnTe(Bi_2Te_3)_n$, which is a single member of a broader class of ternary chalcogenides [32], spans the $n = 0$ phase of MnTe to the $n = \infty$ phase of $Bi_2Te_3$. Between these end members, the structure is composed of interleaving Te-Bi-Te-Bi-Te ($Bi_2Te_3$) quintuple layers (QLs) and Te-Bi-Te-Mn-Te-Bi-Te ($MnBi_2Te_4$) septuple layers (SLs). For example, the $n = 1$ phase is $MnBi_2Te_4$ composed only of SL and the $n = 2$ phase is $MnBi_4Te_7$ composed of alternating SL and QL. The ability to synthesize well-controlled $MnTe(Bi_2Te_3)_n$ thin films will enable tuning the magnetic ground state, and thus the topological phase; it would then be possible to manipulate functional responses of these coupled behaviors by designing heterostructures with other magnets or by controlling the thickness to be predominately even/odd numbers of layers (for an A-type antiferromagnetic odd numbers of layers have a net moment).

However, in contrast to many materials, synthesis of thin films of $MnTe(Bi_2Te_3)_n$ has been confined to the $n = 1$ phase, which requires alternate depositions of MnTe and $Bi_2Te_3$ that, in turn, relies on mixing to form the SL structure [34]. Molecular beam epitaxy (MBE) is well-known as the gold-standard for thin film growth and has been an enabling technology in the study of quantum transport phenomena in topological systems [22,35–37]. The caveat, however, is that for binaries and ternaries the growth of the best quality materials occurs when one or more elements are volatile and can be supplied in excess. Otherwise the film can accumulate errors in stoichiometry as defects, which can easily be in the range of 1-10%. For the case of the $MnTe(Bi_2Te_3)_n$ series, it is conceivable that it can be grown with an adsorption-controlled growth window since both Bi and Te are relatively volatile. However, it is not clear if this is possible since the thermodynamic barrier to intergrowths (due to the layered nature) and antisite defects among Mn and Bi are expected to be relatively low [38]. Here, we report the growth of $MnTe(Bi_2Te_3)_n$ by MBE. For the $n = 1$ phase ($MnBi_2Te_4$), it is shown that the growth is adsorption-controlled based on the Mn flux, where Bi and Te are supplied in excess, though narrow in both temperature and flux ratios. At higher fluxes the $MnBi_4Te_7$ is stabilized. The films properties are in good agreement with data from stoichiometric bulk crystals from the perspective of X-ray diffraction (XRD), Rutherford backscattering spectroscopy (RBS) and magnetotransport. Further, for Mn-rich films clear ferromagnetism emerges, which, together with the $MnBi_2Te_4$ structure, indicates Mn-doped $MnBi_2Te_4$. This work opens many new routes for tuning magnetic topological phases through dimensionality and interfacial phenomena such as the quantized anomalous Hall effect, as well as establishing a route to grow layered ternary chalcogenide materials.

MnTe(Bi$_2$Te$_3$)$_n$ films were grown in a home-built MBE system with base pressure of <5×10$^{-10}$ Torr. The substrates used were (0001) Al$_2$O$_3$, which were heated to 600 °C in the MBE chamber and annealed for 10 minutes prior to growth to clean the surface. Further, high purity Bi, Mn, and Te were calibrated with an *in-situ* quartz crystal microbalance (QCM) to achieve the desired flux ratio. QCM measurements were cross calibrated against film thicknesses measured using X-ray reflectivity, which yielded a flux ratio accuracy of a few percent. Mn flux was kept constant at $1.40 \times 10^{13}$ cm$^{-2}$s$^{-1}$, and the Bi and Te fluxes were set relative to this. Following this, the substrates were cooled to 135 °C where 3 SL were deposited to aid in nucleation. During this step the cells were calibrated to supply the stoichiometric flux of Bi:Mn:Te = 1:2:4, to ensure all films had similar starting growth surfaces. The films were subsequently ramped to the upper growth temperature $185 < T < 250$ °C, where the final depositions took place. The Bi and Te cells were adjusted to supply the appropriate Bi:Mn flux ratio for the sample and a Te overpressure of 2-3 times the stoichiometric value to prevent Te deficiencies in the film. Following the deposition, the films were annealed at the growth temperature for 60 minutes with the Te shutter open.

To understand the structural evolution of MnTe(Bi$_2$Te$_3$)$_n$ films as a function of the Bi:Mn flux ratio as well as the deposition temperature, XRD was performed on two series of samples using Cu-k$_\alpha$ X-rays. Fig. 2 shows the results for XRD $2\theta$-$\theta$ scans where the growth temperature was varied from 185 to 250 °C for the flux fixed at Bi:Mn = 2.3, and the Bi:Mn flux ratio was varied from 1.0-5.0 with the temperature fixed at 225 °C. The data in Fig. 2(a) shows the temperature dependence with the flux in the range to form MnBi$_2$Te$_4$ (thickness of 7 ± 1nm). For films grown at the lowest temperature, 185 °C, the peaks that are resolved are consistent with the (003$n$) series of Bi$_2$Te$_3$, where *n* is an integer. These peaks are marked as solid blue circles. This result implies that at this low temperature Mn is incorporated on the Bi-site and there is insufficient kinetics to form MnBi$_2$Te$_4$. Further, increasing the growth temperature to 200 °C yielded new peaks which are well separated from Bi$_2$Te$_3$ peaks, specifically, around $2\theta$ = 14, 35, and 40°. Upon further increasing the temperature to 225 °C, these peaks became better resolved, and the intensity increased. This series of peaks matches very well to the (003$n$) series of the MnBi$_2$Te$_4$ phase. Upon increasing the temperature to 250 °C, additional peaks emerge around $2\theta$ = 25° and 55°. These are consistent with the formation of MnTe due to the desorption of volatile Bi-Te from the growing surface. Together this implies that all the Mn is adsorbed and the volatility of Bi and Te are controlling the phase formation.

To understand how the Bi-Te volatility controls the growth, Fig. 2(b) shows the detailed dependence of the $2\theta$-$\theta$ scans across a range of Bi:Mn fluxes with the temperature fixed at 225 °C (the thickness of the samples are as follows with increasing Bi:Mn flux ratio - 15, 21, 23, 24, 29, 34, 41, 43, 48 ± 1 nm). Post growth *in-situ* reflection high-energy electron diffraction and *ex-situ* atomic force microscopy images, given in Fig. S1 and S2 in Ref. [39] show the samples have flat 2D surfaces and are highly crystalline. Consider first 2.0 < Bi:Mn < 2.6, which are shown in blue in Fig. 2(b). These curves exhibit the (003$n$) peaks of MnBi$_2$Te$_4$ and the overall character is independent of the incident flux ratio. As discussed more below in the context of RBS measurements and transport, this suggests a growth window for MnBi$_2$Te$_4$. With increasing flux ratio beyond Bi:Mn = 2.6, the peaks appear to split as the next member of the series begins to form. This is highlighted in Fig. 2(b) by the vertical dashed arrows. The new pattern exactly matches MnBi$_4$Te$_7$, indicated as open green triangles. The larger number of peaks is due to the larger unit cell height, which is increased by the addition of a QL unit in between the SLs, as shown in Fig. 1. For the two highest flux values shown (4.5 and 5.0) the peaks are nominally independent of flux ratio, which imply that a growth window may exist for this member. However, given the layered structure, precise homogeneous ordering is unlikely and will require more in-depth study at these higher fluxes. Finally, at fluxes below Bi:Mn = 2.0, shown in red, the MnBi$_2$Te$_4$ phase clearly persists even down to the Bi:Mn = 1.0. This result is very surprising because, with excess Mn, the formation of Mn$_{Bi}$ antisite defects (Mn on the Bi sites), MnTe, or a combination would be expected [38]. However, for the latter there is no sign of MnTe peaks in this measurement.

Although XRD scans indicate that there are windows in the Bi:Mn flux ratio where the different members form, this is not a direct measure of the film stoichiometry. More specifically, non-stoichiometry can be accommodated by the formation of defects while maintaining the crystalline phase. To show that the stoichiometry is indeed controlled, RBS measurements were performed, which directly probe the film stoichiometry to an error less than a few percent. The RBS results are shown in Fig. 3. RBS was performed at the ion beam materials laboratory [40] with 2 MeV He beam at incident, exit and scattering angles of 45º, 70º and 155º, respectively. The He ion energy and large incident/exit angles were used to improve the depth resolution and reduce the uncertainty in determining the elemental concentration of the thin films. Sample spectra are shown in Figs. 3(a) and (b), where the data are solid square symbols and the model fit is the solid red line. At low channel number the broad plateaus are oxygen (onset at channel ≈ 600) and Al (channel ≈ 1000) from the substrate, where the plateau-like peak shape indicates a continuous loss of energy as the He ions penetrate deep into the crystal. With increasing channel number, the additional peaks are Mn (channel ≈ 1400), Te (channel ≈ 1600) and Bi (channel ≈ 1700).

To extract the areal density of each individual species the RBS spectra were fit using the SIMNRA program [41]. Good fits, as shown in Figs. 3(a) and (b), were obtained for all the samples evaluated by RBS. The assumption of a thin film with a fixed stoichiometry used in SIMNRA fitting indicates a uniform film deposition and sharp interface. The results of the fitting are plotted versus the Bi:Mn flux onto the film surface in Fig. 3(c). This figure shows the atomic percentage, $P$(%), of Bi and Mn (the solid data points). Also shown are dashed lines that indicate the ideal atomic percentage for MnBi$_2$Te$_4$ and MnBi$_4$Te$_7$. For 2.0 < Bi:Mn < 2.6, the atomic percentage of the film is found to be very close to the ideal value expected for MnBi$_2$Te$_4$. This supports the analysis from XRD, which showed peaks consistent with this phase. The fact that the film stoichiometry is nominally the same with increasing flux of Bi implies that the excess Bi and Te are desorbed off the growing surface rather than incorporated. Further, the flux ratio of Bi:Mn = 4.5 similarly shows that the film stoichiometry agrees well with MnBi$_4$Te$_7$. Finally, on the lower end of Bi:Mn, the stoichiometry is found to be roughly 1:1. This is very surprising since the predominate phase found in XRD is that of MnBi$_2$Te$_4$, without features of MnTe. This implies that there may be a large stability window for the formation of MnBi$_2$Te$_4$ despite high levels of non-stoichiometry.

Figs. 4 (a) and (b) show detailed temperature dependent transport data for the films grown at 225 °C from 300 K down to 2 K for Bi:Mn = 2.0 and 5.0, respectively. Here, this data is shown in sheet resistance $R(\Omega/\square)$ rather than resistivity ($\rho = R(\Omega/\square) \times thickness$) due to the convolution of the bulk states and surface states. Both samples show metallic behavior, where the resistance decreases with decreasing temperature. The resistance for MnBi$_2$Te$_4$ films is found to be significantly larger than MnBi$_4$Te$_7$. This can be attributed to the structure of MnBi$_4$Te$_7$ which is composed of interleaved SL and QL where the additional layer of Bi$_2$Te$_3$ likely provides higher conductivity pathways. Further, this coincides with transport data from bulk crystals where the low temperature resistivity of the MnBi$_2$Te$_4$ is around 1 mΩcm and MnBi$_4$Te$_7$ is around 0.4 mΩcm [42,43], which agrees well with 1.63 mΩcm for the MnBi$_2$Te$_4$ film and 0.55 mΩcm for the MnBi$_4$Te$_7$ film. Finally, the magnetic transitions can be clearly identified in Figs. 4(a) and (b) as kinks.

These are highlighted in the insets where the kinks occur roughly at 25 K for MnBi$_2$Te$_4$ and 11 K for MnBi$_4$Te$_7$, which, again, agree well with the bulk.

Figs. 5(a) and (b) show the magnetic field dependence of the transport at 3 K for various Bi:Mn flux ratios where the samples were measured in van der Pauw geometry. Here the magnetic field, $H$, is applied in the out-of-plane direction along the crystallographic c-axis. Shown in Fig. 5(a) is the rescaled magnetoresistance $R_N = (R - R_{Min.})/(R_{Max.} - R_{Min.})$, where $R_{Min.}$ and $R_{Max.}$ are the maximum and minimum resistance, respectively, which have been offset for clarity. Plotting in this way enables a direct comparison of the shape and field dependence across the full range in fluxes. For the flux ranges of Bi:Mn = 1.0-3.2 the curves are found to be nominally the same, independent of the flux ratio. These are found to be quite flat between the applied fields of -2 T to 2 T. At |H| > 3.0 T, $R_N$ turns upward with increasing field and peaks at a local maximum. Beyond this field $R_N$ decreases until around 8 T where there is an upturn and $R_N$ increases with increasing field. The local maximum in $R_N$ coincides with the behavior in the Hall effect versus flux ratio which shows a kink that occurs in the same field range, see Fig. 5(b). In comparison to bulk data, this can be associated with the spin-flop phase transition where the spins on the Mn sites, initially oriented along the c-axis, flip 90° and point in-plane [43]. For the sample with Bi:Mn = 5.0, $R_N$ is found to be dramatically different. At low-field this sample shows a slight negative magnetoresistance with a small hysteresis around $|H| < 0.2$ T. This is consistent with the observations in bulk crystals. Finally, in comparison to bulk samples for MnBi$_2$Te$_4$, while the magnetic ordering temperature and magnetic field scales for the spin-flip transition is similar to bulk samples, the overall curve shape differs. In bulk crystals and exfoliated flakes, the magnetoresistance is nominally flat at low magnetic fields and undergoes a steep drop at the spin-flop transition and a weak peak at the spin-flip transition around 8 T [43–45]. However, the shape of the magnetoresistance is found to match closely to that of exfoliated flakes where the carrier concentration has been tuned with an external gate electrode, with virtually no peak at higher field that indicates the spin-flip transition [26]. It is noted that the overall resistance value observed here is significantly lower.

The Hall resistance $R_{xy}$, shown in Fig. 5(b) gives additional insight into the evolution of the magnetic phase and the defect density with Bi:Mn that compliments $R_N$. The overall trends can be more easily seen in Fig. 5(c) where a linear slope has been subtracted from the Hall resistance ($R_{xy}$-$R_{xy,linear}$) and offset for clarity. First, for Bi:Mn = 2.0 and 2.2 the curves are nominally identical. This supports the XRD and RBS analysis for a narrow growth window for MnBi$_2$Te$_4$. This agrees well with data from bulk samples where the Hall resistance shows a steep drop at the spin-flop transition around $H = 3.5$ T, and slight change in slope around 6-8 T. These transitions can clearly be seen in the plot of $R_{xy}$-$R_{xy,linear}$. The spin-flop transition is highlighted by vertical black dashed lines, which is nominally independent of flux, consistent with the dependence of $R_N$. Further, the upturn at around 6-8 T likely indicates the spin-flip transition. Both these transitions can clearly be resolved between Bi:Mn = 1.8-3.2. Moving to the highest flux ratio Bi:Mn= 5.0, $R_{xy}$-$R_{xy,linear}$ shows only a slight hysteresis around $|H| < 0.2$ T. This is consistent with the hysteresis found in $R_N$ in Fig. 5(a) and agrees well with bulk data for MnBi$_4$Te$_7$. Interestingly, the samples with Bi:Mn = 2.6 and 3.2 show a low-field hysteresis that agrees with MnBi$_4$Te$_7$, highlighted by vertical red dashed lines. The low-field behavior of MnBi$_4$Te$_7$ is superimposed on the spin-flop and spin-flip characteristic of MnBi$_2$Te$_4$, but also a slightly larger field hysteresis that closes at about 3 T, marked by vertical blue lines. This implies that within this intermediate regime there is a phase mixture, which is overall consistent with the XRD. Lastly, at the lowest flux ratio Bi:Mn = 1.0, a large hysteresis emerges, which is marked by vertical light blue dashed lines. Since the XRD indicates that the phase is MnBi$_2$Te$_4$, this likely implies that this ferromagnetic phase is due to Mn-doping of MnBi$_2$Te$_4$; similarly, for Bi:Mn = 2.6 and 3.2, this may also explain the hysteresis that closes at around 2 T as Mn-doped MnBi$_4$Te$_7$.

Finally, shown in Fig. 5(d) is the carrier density, $N_{3D}$, extracted from the Hall effect at 150 K, well above the magnetic transition temperature, as shown in Fig. S3 of Ref. [39]. $N_{3D}$ can be separated into 2 different regimes. In the first regime, for Bi:Mn < 4, the structure extracted from XRD is MnBi$_2$Te$_4$, and $N_{3d}$ shows a small increase with increasing Bi:Mn at low values, then a slight plateau just above Bi:Mn = 2. In the second regime, for values above Bi:Mn = 4, there is a significant increase in carrier concentration. This coincides with the formation of MnBi$_4$Te$_7$ from XRD. In both regimes the carrier type is always found to be electron-like. The data is consistent with the expectation from bulk samples where the carrier concentration at room temperature is of the order of $1\times10^{20}$ cm$^{-3}$ for MnBi$_2$Te$_4$ [43] and $6\times10^{20}$ cm$^{-3}$ for MnBi$_4$Te$_7$ [42]. At these high densities the Fermi level is well into the conduction band. In comparison to the $n = \infty$ members such as Bi$_2$Te$_3$, carrier densities can be much lower than observed here. Further, in both of these materials, antisite type $Bi_{Mn}^+$ and $Mn_{Bi}^-$ defects will form to compensate nonstoichiometries as vacancies and interstitials are too high in energy [38]. Excess Bi will lead to the formation of $Bi_{Mn}^+$ donors which will raise $N_{3d}$; while Mn excess creates $Mn_{Bi}^-$ acceptors, which will lower $N_{3D}$ and move the Fermi level closer to the conduction band minimum. This is reflected in the data with the trend of increased carrier concentration as the Bi:Mn ratio is increased in the MnBi$_2$Te$_4$, and may also explain how the samples are able to remain in the MnBi$_2$Te$_4$ structure at Bi:Mn < 2. This highlights the challenge in synthesizing these more complex structures of MnBi$_2$Te$_4$ and MnBi$_4$Te$_7$, where charged antisite defects occur with large density and are highly dependent on film stoichiometry [38]. Finding suitable conditions that minimize such defects while maintaining phase purity represents the next challenge for this class of materials. MBE growth, however, offers several routes to tune the global Fermi level. This includes utilizing *in-situ* virtual substrate methods that reduce interfacial defects [36] and alloying with MnSb$_2$Te$_4$ to balance acceptor and donor antisite defects where the n-type to p-type crossover occurs around $x = 0.63$ in Mn(Bi$_{1-x}$Sb$_x$)$_2$Te$_4$ [46].

To conclude, we have shown the growth of MnTe(Bi$_2$Te$_3$)$_n$ by MBE. This utilizes a two-step growth sequence where a low temperature layer was deposited with stoichiometric flux ratio, which was followed by a higher temperature deposition where excess Bi and Te evaporate. Using this method, the $n=1$ (MnBi$_2$Te$_4$) and $n=2$ (MnBi$_4$Te$_7$) can be stabilized. Additionally, we have shown the MnBi$_2$Te$_4$ phase can be grown in excess Bi-Te flux. However, the growth window is relatively narrow in both temperature and flux. The resulting films are found to be single phase by XRD and stoichiometric by RBS. Moreover, transport properties are in good agreement with bulk crystals of MnBi$_2$Te$_4$ and MnBi$_4$Te$_7$ with magnetic transitions at around 25 K and 11 K, respectively. Understanding how to synthesize this class of materials by MBE opens a huge space for engineering magnetism and topological states in a single material without the need for doping other elements. Moreover, this can enable functional tuning by integration as heterostructures with ferromagnetic and antiferromagnetic materials, both of which can exhibit a wide variety of transition temperatures, moments, orderings, and domain structures. These together can enable realization of exotic phases or raise the temperature where such phases can be observed, with a particular emphasis on the QAH phase. Towards this goal many challenges remain. Of key importance is controlling the Fermi level, which requires a deeper understanding of what defects form and under what conditions. Achieving such goals will open many new routes to control this novel material family and guide stabilizing other exotic systems as high-quality thin films by MBE.

## Acknowledgments

This work was supported by the U.S. Department of Energy, Office of Science, Basic Energy Sciences, Materials Sciences and Engineering Division (MBE growth, transport, and X-ray diffraction), and by the Laboratory Directed Research and Development Program of Oak Ridge National Laboratory, managed by UT-Battelle, LLC, for the U. S. Department of Energy (part of MBE growth). Ion beam analysis was performed at the Ion Beam Materials Laboratory located at the University of Tennessee, Knoxville. YZ acknowledges support from Energy Dissipation to Defect Evolution (EDDE), an Energy Frontier Research Center funded by the U.S. Department of Energy, Office of Science, Basic Energy Sciences, under contract number DE-AC05-00OR22725 (RBS).

We would like to thank Jiaqiang Yan for insightful discussions.## References

**Figures**

**Figure 1**

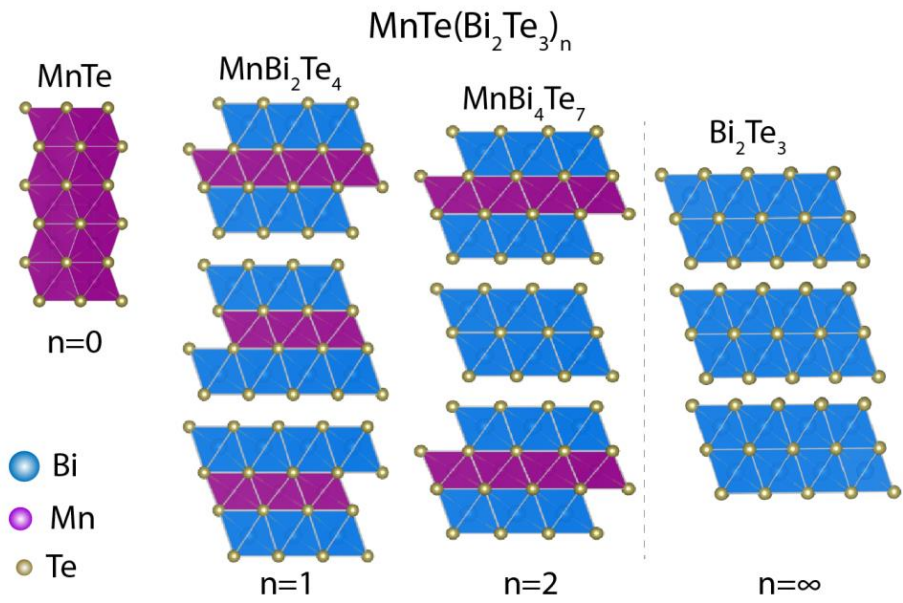

**Figure 1**. Schematic crystal structures of the MnTe(Bi$_2$Te$_3$)$_n$ system for various *n*.

**Figure 2**

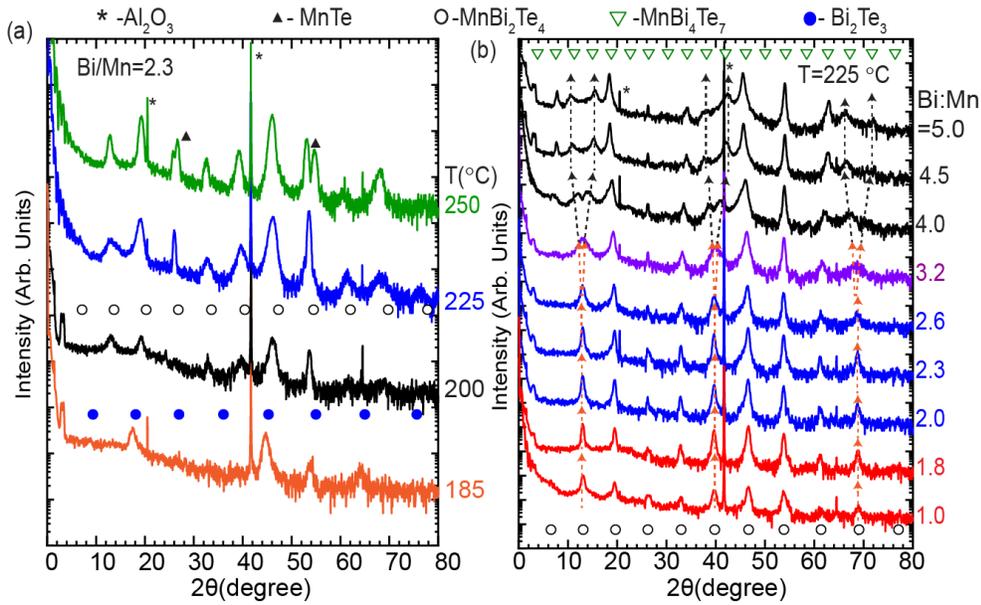

**Figure 2**. X-ray diffraction 2θ scans for (a) Bi:Mn flux ratio fixed at 2.3 at various growth temperatures and (b) temperature fixed at 225 °C for various Bi:Mn fluxes, with the exact flux ratio indicated on the right hand axis. As indicated the peaks are marked as astericks-$Al_2O_3$, solid triangles-MnTe, open circles-$MnBi_2Te_4$, open triangles-$MnBi_4Te_7$, and solid circles-$Bi_2Te_3$. Dashed vertical arrows highlight the emergence of $MnBi_4Te_7$ in (b).

**Figure 3**

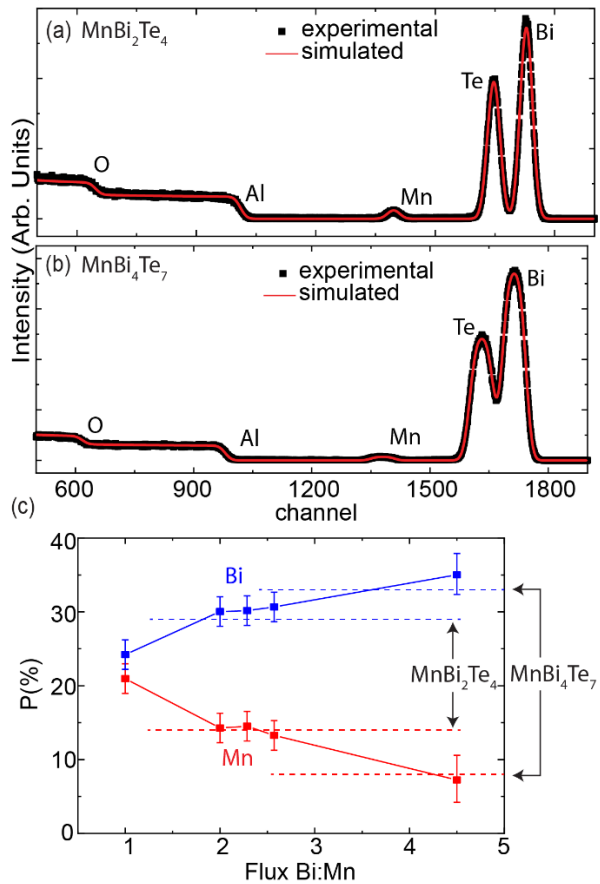

**Figure 3**. (a-b) RBS for two samples with flux ratio Bi:Mn = 2 (a) and 4.5 (b). (c) Results of fitting the spectra versus the flux incident on the sample where $P(\%)$ is the atomic percentage of Mn (red) and Bi (blue). The horizontal dashed lines indicate the atomic percentages expected for $MnBi_2Te_4$ and $MnBi_4Te_7$.

**Figure 4**

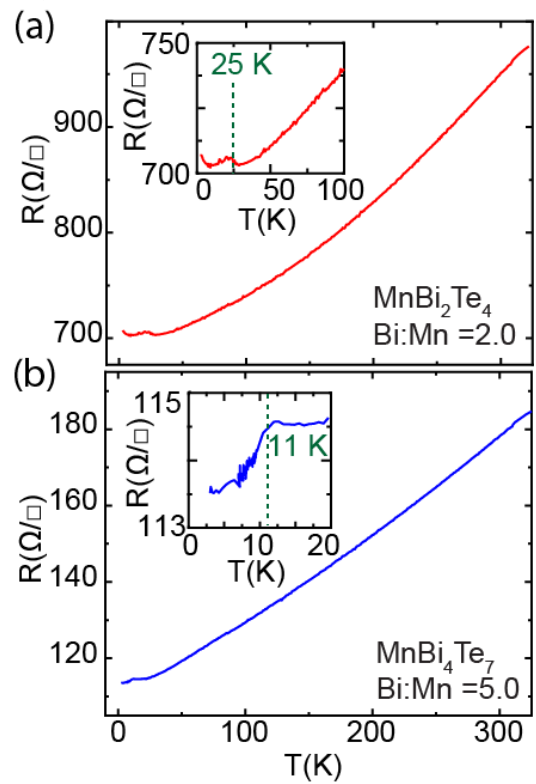

**Figure 4** (a-b) Resistance versus temperature plots for MnBi$_2$Te$_4$ with Bi:Mn = 2.0 (a) and MnBi$_4$Te$_7$ with Bi:Mn = 5.0 (b). The insets highlight the low temperature transport where the kinks indicate the magnetic transition, highlighted by the vertical dashed lines.

**Figure 5**

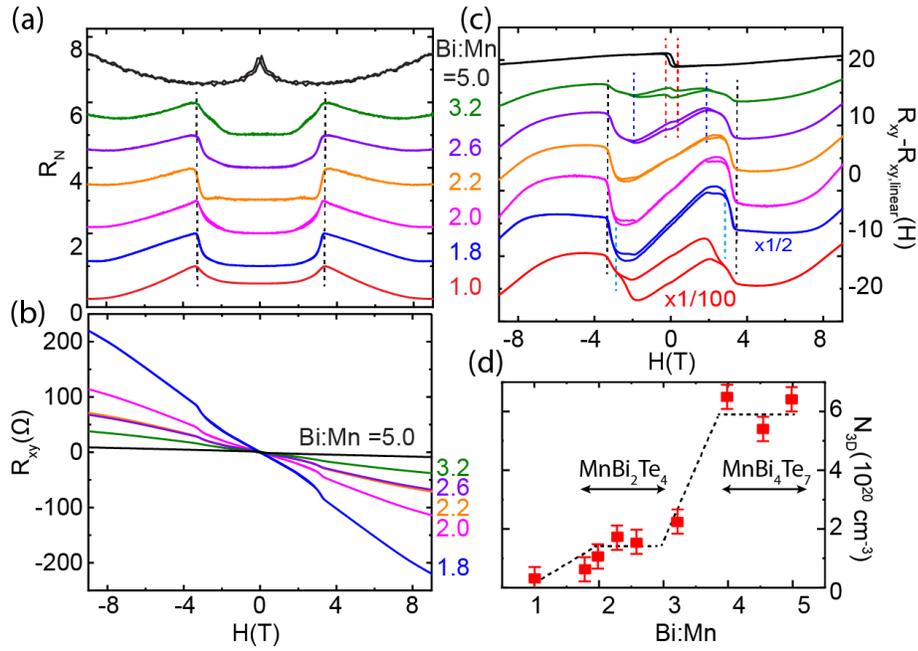

**Figure 5**
(a-b) Normalized magnetoresistance $R_N = (R_S - R_{S,Min.})/(R_{S,Max.} - R_{S,Min.})$ (a) and Hall resistance $R_{xy}$ (b) versus magnetic field for various Bi:Mn flux ratios taken at 3 K. (c) Hall resistance in (b) with the linear portion subtracted. Vertical dashed lines indicate magnetic transitions, see main text for details and discussion. In (a-c) Bi:Mn = 1.0-2.6 exhibit the MnBi$_2$Te$_4$ phase while, Bi:Mn>3.2 exhibits MnBi$_4$Te$_7$. (d) Carrier concentration extracted from the Hall effect at 150 K versus Bi:Mn ratio.

# Supplemental Materials:

# Adsorption-controlled growth of MnTe(Bi$_2$Te$_3$)$_n$ by molecular beam epitaxy exhibiting stoichiometry-controlled magnetism

Jason Lapano[1], Lauren Nuckols[2], Alessandro R. Mazza[1], Yun-Yi Pai[1], Jie Zhang[1], Ben Lawrie[1], Rob G. Moore[1], Gyula Eres[1], Ho Nyung Lee[1], Mao-Hua Du[1], T. Zac Ward[1], Joon Sue Lee[3], William J. Weber[2], Yanwen Zhang[1,2], Matthew Brahlek[1]*

[1]Materials Science and Technology Division, Oak Ridge National Laboratory, Oak Ridge, TN, 37831, USA
[2]Department of Materials Science and Engineering, University of Tennessee, Knoxville, Tennessee 37996, USA
[3]Department of Physics and Astronomy, University of Tennessee, Knoxville, Tennessee 37996, USA

Correspondence should be addressed to *brahlekm@ornl.gov

**Fig. S1.** Reflection high energy electron diffraction (RHEED) images taken for films grown at various Bi:Mn flux ratios. Images were taken between 100-150°C after cooling from the growth temperature in the molecular beam epitaxy chamber. All films showed streaky diffraction spots, indicative of smooth, uniform films regardless of stoichiometry.

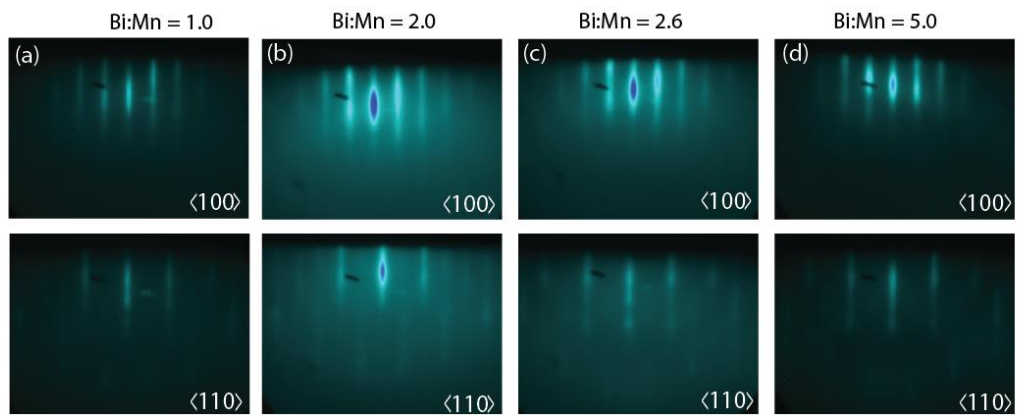

**Fig. S2.** Atomic force microscopy images of MnTe(Bi$_2$Te$_3$)$_x$ thin films showing evolution of surface morphology for Bi:Mn ratios of 1 (a), 2 (b), 2.6 (c) and 5 (d). Small pillars are visible in (a) only for low Bi:Mn ratios, and are not observed at higher fluxes (b-d). Scale bars are 1 μm.

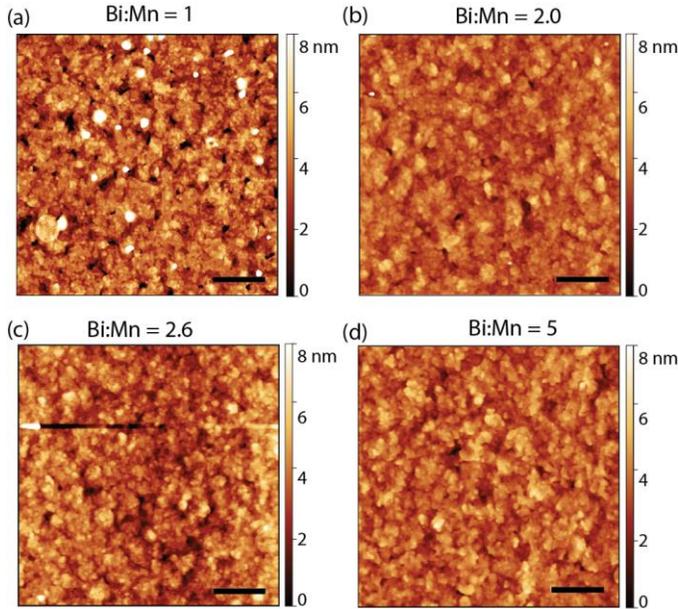

**Fig. S3.** Hall resistance of samples as a function of Bi:Mn flux ratio. Results are antisymmetrized to remove the magnetoresistance component, and measurement is taken at 150 K, well above T$_N$ to avoid magnetic contributions. R$_H$ is used to obtain the 2D sheet carrier density, which is converted to 3D carrier density from the extracted film thickness and is plotted in Fig. 5(d).

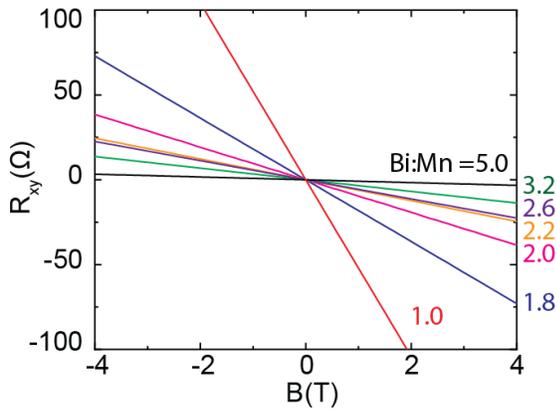